\documentclass[twoside,twocolumn,a4paper,10pt]{revtex4-1}

\usepackage{graphicx}
\usepackage[english]{babel}
\usepackage[T1]{fontenc}
\usepackage[usenames,dvipsnames]{xcolor}

\usepackage{hyperref}

\usepackage{amsmath,mathtools}

\iftrue
  \def\mybox#1{#1}
  \def\incg#1{\includegraphics{#1}}
\else
  \def\mybox#1{\rule{\linewidth}{2.5cm}}
  \def\incg#1{\rule{1ex}{1ex}}
\fi

\newcommand\fref[1]{Fig.~\ref{#1}}
\newcommand\eref[1]{Eq.~\eqref{#1}}

\newcommand{\ie}{\emph{i.\,e.}}
\newcommand{\etal}{\emph{et al.}}

\newcommand{\Gf}{\mathbf G}
\newcommand{\Ham}{\mathbf H}
\newcommand{\Over}{\mathbf S}
\newcommand{\SE}{\boldsymbol\Sigma}
\newcommand{\Gam}{\boldsymbol\Gamma}
\newcommand{\Spec}{\mathbf A}

\newcommand\unit[2]{#1\,#2}
\newcommand\unitr[2]{\unit{#1}{\mathrm{#2}}}
\newcommand\Ang{\text\AA}
\newcommand\Cm{\mathrm{cm}}

\newcommand\subref[1]{#1)}

\begin{document}

\title{Manipulating the voltage drop in graphene nanojunctions using a gate potential}

\author{Nick R. Papior}
\email{nickpapior@gmail.com}
\author{Tue Gunst}
\author{Daniele Stradi}
\author{Mads Brandbyge}
\affiliation{Center for Nanostructured Graphene (CNG), Denmark}
\affiliation{Department of Nanotech, Technical University of Denmark, H.C. \O rsteds plads
345b, Kgs. Lyngby, Denmark}

\begin{abstract}
  Graphene is an attractive electrode material to contact nanostructures down to the
  molecular scale since it can be gated electrostatically. Gating can be used to control
  the doping and the energy level alignment in the nanojunction, thereby influencing its
  conductance.  Here we investigate the impact of electrostatic gating in nanojunctions
  between graphene electrodes operating at finite bias. Using first principles quantum
  transport simulations, we show that the voltage drop across \emph{symmetric} junctions
  changes dramatically and controllably in gated systems compared to non-gated junctions. In
  particular, for \emph{p}-type(\emph{n}-type) carriers the voltage drop is located 
  close to the electrode with positive(negative) polarity, \ie\ the potential of the junction 
  is pinned to the negative(positive) electrode. We trace this
  behaviour back to the vanishing density of states of graphene in the proximity of the
  Dirac point. Due to the electrostatic gating, each electrode exposes different density of
  states in the bias window between the two different electrode Fermi energies, 
  thereby leading to a non-symmetry in the voltage drop across
  the device. This selective pinning is found to be independent of device length when 
  carriers are induced either by the gate or dopant atoms, indicating a general effect for electronic circuitry based
  on graphene electrodes. We envision this could be used to control the
  spatial distribution of Joule heating in graphene nanostructures, and possibly the
  chemical reaction rate around high potential gradients.
\end{abstract}

\maketitle


\section{Introduction}

Graphene (Gr) shows great promise as a central material for future two-dimensional (2D)
nanoelectronic applications\cite{Geim2007,Raza2011}. In particular, its semi-metallic
character and its record high mean-free path\cite{Bolotin2008} make it a top candidate for
ultra-fast and flexible electronic components\cite{Kim2009,Georgiou2013}. Fuelled by these
perspectives, nanostructured devices down to the molecular scale using electrodes based on
Gr have recently been put forward\cite{Cao2013,Jia2013,Ullmann2015,Sadeghi2015}. In their
most generic form, these devices are composed by a Gr constriction where the
narrowest junction consists of a Gr nanoribbon (GNR)\cite{Dutta2010,Tang2013} or an organic
molecule\cite{Cao2013,Jia2013,Ullmann2015,Sadeghi2015}. More complex structures such as
Gr antidot lattices\cite{pedersen_graphene_2008,Bai2010} can also be viewed as
consisting of a network of constrictions.

A unique feature of Gr electrodes
is that their electronic properties can easily be tuned by electrostatic gating. In fact,
electrostatic gates can be used to increase the carrier density in Gr up to above
$\unit{10^{13}}{\Cm^{-2}}$.\cite{Ye2011} For ion gating it has even been
    possible to reach carrier densities of $\unit{10^{14}}{\Cm^{-2}}$ which corresponds to
    a Fermi energy shift of about $\unitr{1}{eV}$.\cite{Yin2014} It has been shown that gating can be used
to tune the resonances localized in the narrowest part of the
junction,\cite{Song2009,Prins2011} as the electronic states of the electrodes are usually
affected only weakly by the gate-induced capacitive field due to effective screening by the high density of states(DOS)\cite{Santos2013}.  
However, for Gr electrodes, the lower DOS and its flat geometry, makes it comparable to the junction itself, and 
is thus likely to be perturbed similarly by gating. This peculiarity leads to a novel, yet largely
unexplored, paradigm for graphene-based electronics, as the transport characteristics of
the device ultimately depend on the response of the entire system to the gate. In
electronic transport simulations, the effect of electrostatic gating and induced doping
charge in the device has often been mimicked by rigidly shifting the position the
Fermi-level/chemical potential in calculations without explicitly including the gate or
dopants\cite{Ryndyk2012,Nikolic2012,Gunst2013,Lu2014}. However, despite accounting for some of the
effects, these approaches completely neglect the self-consistent response of the device to 
the additional charge doping or the gate-induced electric field.

\begin{figure}
  \centering
  \mybox{\includegraphics{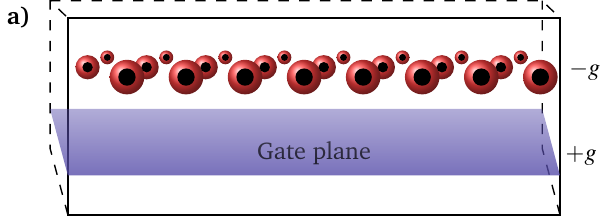}}
  \vskip 4pt
  \mybox{\includegraphics{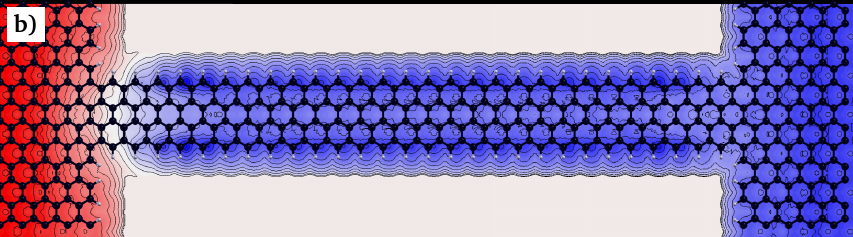}}
  \caption{\subref a implementation of the field effect gate model. Redistribution of charge
      from atoms to gate plane. \subref b resulting voltage drop for a $\unitr{8.3}{nm}$
      long constriction including a field effect gate of
      $n=2\times10^{13}e^-/\mathrm{cm}^2$ and a bias of $\unitr{0.5}V$. Contour lines are
      separated by $\unitr{0.022}V$.
      \label{fig:impl-6x}
  }
\end{figure}

Here, we investigate these issues by extending the TranSIESTA first principles electronic
transport package, based on density functional theory and nonequilibrium Green function
(DFT-NEGF)\cite{Soler2002,Brandbyge2002}, with the inclusion of a physically motivated
gate model, see \fref{fig:impl-6x}a. These improvements allow us to consider on an equal
footing the effect of charge doping, capacitive gate field, and of the finite bias voltage
in our DFT+NEGF simulations (see the Methods section for additional details of the
implementation). We apply this methodology to Gr constrictions consisting of nanoribbon
junctions between Gr electrodes. For these we demonstrate how the transport
characteristics depend in a non-trivial way on the applied source/drain and gate
voltages. As seen in \fref{fig:impl-6x}b, upon gating and bias, the voltage drop is pinned
to the electrode of a given polarity depending on the doping type and the bias, even for a
constriction of $\unitr{8.3}{nm}$. %
We can relate the phenomenon to the gate-dependent behavior
of the voltage drop in the system which, in turn, can be traced back to the energy
dependence of the DOS in the Gr electrodes. The electronic structure
of the semi-metallic Gr electrodes display zero DOS at its charge neutrality point, and a
linear increase of the DOS away from it (V-shape). Our analysis demonstrate how the
V-shaped DOS in the electrodes controls the voltage drop in the Gr junctions indicating a
quite generic scenario.

The control of the position of the voltage drop on the nano-scale with gate could be
useful in practical applications. We envision this feature, f.ex., could be used to tune the
spatial distribution of Joule heating in the device and influence its breaking at the
nanoscale\cite{Murali2009,Gunst2013}. Our results highlight the importance of using fully
self-consistent electronic transport simulations to predict and design the gating behavior
under operating conditions of the emerging class of devices with electrodes having a
vanishing DOS\cite{Santos2013}.

\section{Results and discussion}
\label{sec:appli}

\begin{figure}
  \centering %
  \mybox{\includegraphics{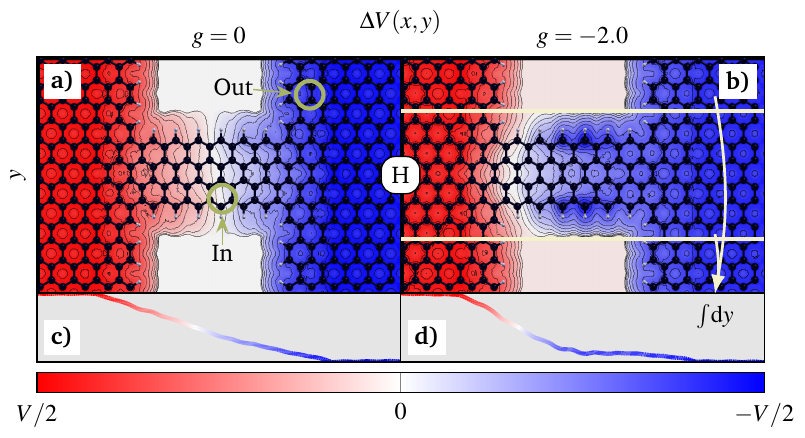}}
  \caption{Electronic Hartree potential drop integrated perpendicular to the plane and
      above a cutoff electron density $\rho_\epsilon=\unit{0.008}{e/\mathrm{\Ang}^{-3}}$
      and projected to the graphene plane for the Hydrogen GNR, \subref a, \subref
      b. \subref c and \subref d are the contour 
      plot further integrated in the box indicated in \subref b. The non-gated system
      shows a linear gradient, whereas for $g<0$ (\emph n-doped) a pinning of
      the potential towards the right (positive) electrode.
      \label{fig:bias-drop-H}
  }
\end{figure}

\begin{figure}
  \centering %
  \mybox{\includegraphics{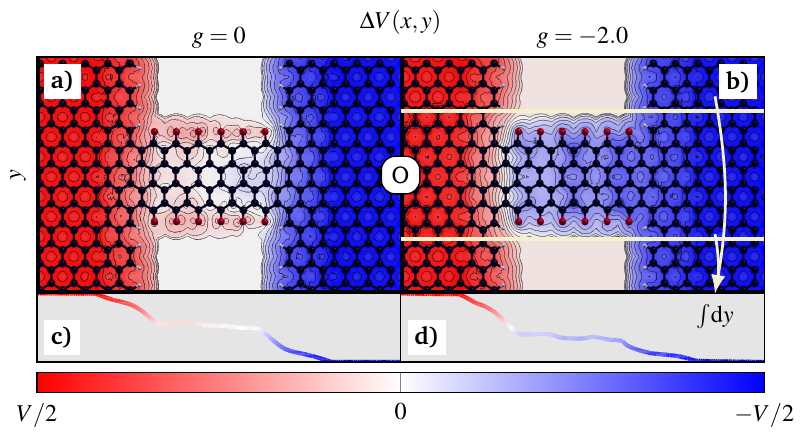}}
  \caption{Electronic Hartree potential drop integrated perpendicular to the plane and
      above a cutoff electron density $\rho_\epsilon=\unit{0.008}{e/\mathrm{\Ang}^{-3}}$
      and projected to the graphene plane for the Oxygen GNR, \subref a, \subref b.
      \subref c and \subref d are the contour
      plot further integrated in the box indicated in \subref b. The non-gated system
      shows a gradient at the GNR boundary, whereas for $g<0$ (\emph n-doped) a pinning of
      the potential towards the right (positive) electrode.
      \label{fig:bias-drop-O}
  }
\end{figure}

We have applied our method to two geometrically similar, ``left-right'' symmetric Gr
nanojunctions, formed by a Gr nanoribbon connected to pristine Gr electrodes, see
\fref{fig:bias-drop-H} and \fref{fig:bias-drop-O}. For zero gate/doping ($g=0$) the former
yield an electron-hole symmetric electronic structure (Hydrogen GNR), whereas the latter
yield a $e$--$h$ non-symmetric electronic structure (Oxygen
GNR)\cite{Cervantes-Sodi2008,Cantele2009,Selli2012}. The hydrogen-terminated system is also
investigated using dopant atoms instead of the electrostatic gating\cite{Tang2013}. The
simulation unit cell has an area of $\sim200$ Gr unit cells. The gate is placed
$\unitr{20}{\Ang}$ beneath the planar Gr structure and we sweep the gating levels ($g$)
according to $g\times\unit{10^{13}}{e^-/\Cm^{2}}$.

In \fref{fig:bias-drop-H}, \subref a and \subref b, we plot the potential drop across the Gr
constriction at $\unitr{0.5}V$ for $g=0$, \subref a, and \emph{n}-doped with $g=-2$
gating, \subref b. The potential profile has been integrated in the perpendicular
direction to the Gr surface for electronic densities above
$\rho_\epsilon=\unit{0.008}{e/\mathrm{\Ang}^{-3}}$ projected onto the $x$--$y$ plane. The
lower panels, \subref c and \subref d, is a further projection onto the transport
direction ($x$) as indicated in \fref{fig:bias-drop-H}b. At $g=0$ we obtain an
anti-symmetric potential drop in the transport direction ($\Delta V(x)=-\Delta V(-x)$) as
expected for a fully $e$--$h$ and left-right symmetric constriction. On the other hand, in
the $g=-2$ ($n$-doped Gr), we see a clear pinning of the potential profile to the positive
electrode, \ie\ the potential drop at the negative electrode. Conversely, calculations
with $g=+2$ ($p$-doping) with $\unitr{0.5}V$ displays a pinning at the negative electrode,
while for $g=+2$ and $\unitr{-0.5}V$ we regain the plot shown. This confirms the geometric
symmetry.

\begin{figure}
  \centering%
  \mybox{\includegraphics{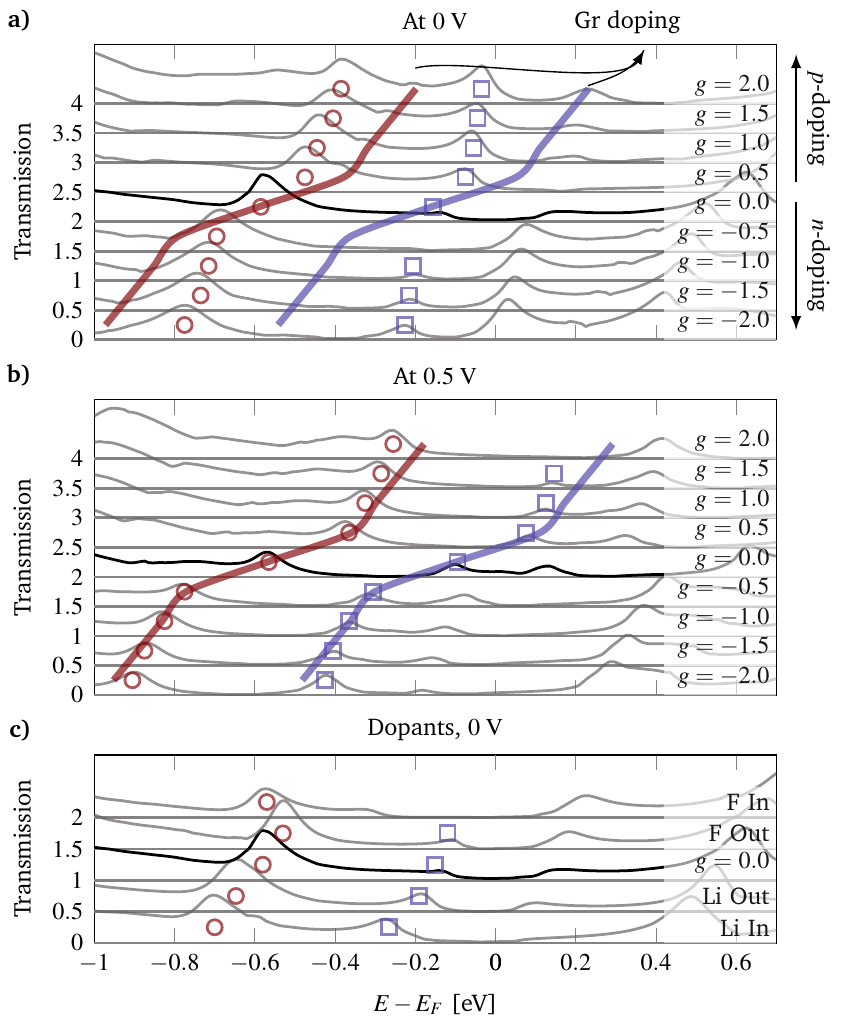}}
  \caption{Transmission spectra for the constriction at various doping levels for
      $\unitr0V$, \subref a, $\unitr{0.5}V$, \subref b, and for dopants, \subref c. The
      middle line (black) at zero gating is a symmetric transmission function with two
      distinct resonances (marked, \protect\incg{square} and \protect\incg{circle}). Gating
      the constriction shifts the resonances as indicated by the displacements of the marks. The full
      lines, crossing vertically the different doping levels, indicates the graphene electrode Fermi level shift due to the doping
      aligned at the $g=0$ mark.
      \label{fig:constriction-both}
  }
\end{figure}

In \fref{fig:constriction-both} we show the electron transmission spectra for the hydrogen
passivated constriction at $\unitr0V$, \subref a, and $\unitr{0.5}V$, \subref b, for
different values of $g$ each vertically shifted $1/2$. As a measure of gating we track the
position of two resonances, \incg{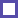} and \incg{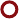} dots, corresponding to a
resonance in the constriction located at the edge and in the center, respectively. The
middle thick line is the transmission for $g=0$, and is equivalent to earlier results
where these resonances are discussed\cite{Gunst2013}. In addition, we plot the energy
shift of the Dirac point for pristine graphene as vertical lines aligned at each of the
two resonances at $g=0$. These vertical lines matches exactly the
shift in chemical potential due to the doping in the electrodes. Discrepancies between the
electrode gating (lines) and the resonance positions (dots) illustrates the difference in
just rigidly shifting the resonances according to electrode doping, and a fully
self-consistent calculation of the resonance positions.  Importantly, at $\unitr0V$ we
find that the resonance peaks does \emph{not} simply follow the gating. Moreover, the two
peaks are shifting/gated independently of each other; the center resonance peak,
\incg{circle}, follows the pristine doping closer than the edge resonance peak,
\incg{square}, due to a difference in electrode coupling between the
resonances. On the other hand, at $\unitr{0.5}V$ we find that both
peaks follow the pristine graphene electrode doping.
As shown in \fref{fig:bias-drop-H}b, the junction behaves as an extension of the positive
electrode and therefore the resonance position is pinned at the Fermi level of this
particular electrode. The self-consistent calculation is needed to capture the correct
transition with bias from semi-independent resonances to the pinned behavior. The same
calculation was performed on a $\unitr{8.3}{nm}$ long ribbon \fref{fig:impl-6x}b
exhibiting the same pinning feature.

\fref{fig:bias-drop-O} are for the Oxygen terminated graphene nanoribbon. This nanoribbon
has no $e$--$h$ electronic DOS
symmetry\cite{Cervantes-Sodi2008,Cantele2009,Selli2012}. Similarly to the Hydrogen system
we calculate for $g=0$ and $g=-2$ at $\unitr{0.5}V$. \subref a shows that the Oxygen edges
pins slightly to the negative electrode for zero gating, while gating, \subref b, the
entire ribbon is pinned to the positive electrode, equivalent to the Hydrogen case
\fref{fig:bias-drop-H}b. This is also seen in the projected potential profiles
\fref{fig:bias-drop-O}c and \subref d. This confirms that the selectivity of the potential
profile in the gated devices does not rely on the $e$--$h$ symmetry of the junction, and
conjectures the generality of this behavior in systems with electrodes having V-shaped DOS
around $E_F$, regardless of the electronic structure of the central part connecting the
two electrodes.

\begin{figure}
  \centering%
  \mybox{\includegraphics{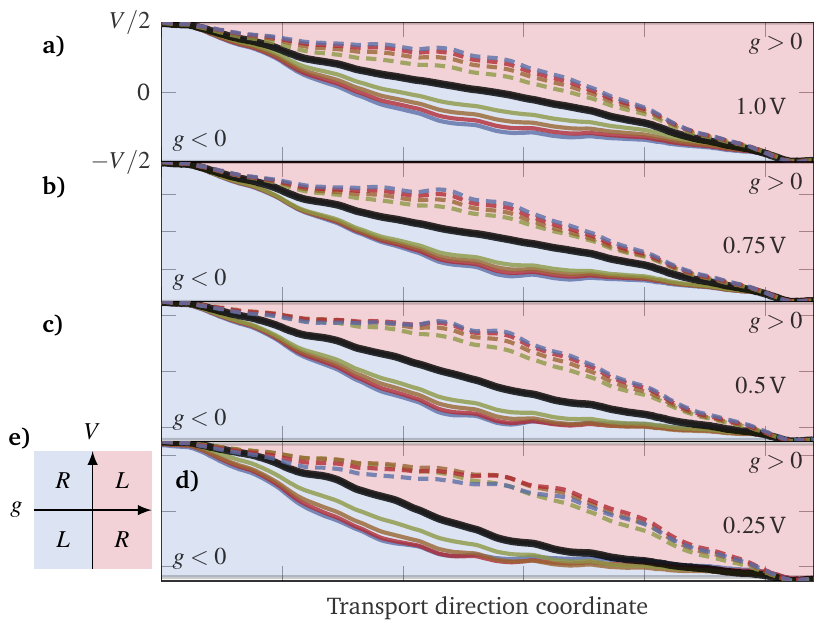}}
  \caption{Integrated Hartree potential profile in a region of width corresponding to the
      ribbon along the entire constriction. The thick middle line is the potential profile for
      $g=0$. The blue regions correspond to \emph{n}-doped
      graphene (full lines), while red are \emph{p}-doped graphene
      (dashed lines). The non-gated calculations show a linear behavior whereas gated systems have a
      non-symmetry between the left and right electrode DOS breaking the left-right anti-symmetry in the potential drop.
      \subref e summarizes the trends where $L$/$R$ means pinning to the
      left/right electrode.
      \label{fig:pot-drop}
  }
\end{figure}

The generic behavior of the potential drop just outlined is summarized in
\fref{fig:pot-drop}, which shows the one-dimensional potential drop calculated for the
hydrogen-terminated constriction for a number of different gates and positive bias
voltages, similar to that of \fref{fig:bias-drop-H}c and \subref d. Independently on the
particular value of the bias voltage applied, gating the system always leads to a marked
asymmetry of the potential drop across the constriction. For any value of \emph{n}-doping,
the potential drop pins always to the positive (right) electrode for positive
bias. Similarly, for any value of \emph{p}-doping, the system couples to the negative
(left) electrode for positive bias. These results further demonstrate the general
phenomenon that does not depend on the particular values of applied gate or bias voltage.
Furthermore, our calculations highlight the important fact that the charge neutrality
point for the electrodes is a special case which does not extrapolate to the gated case.
This becomes even more important if one considers the experimental difficulties in
retaining a charge neutral sample\cite{Martin2007,newaz_probing_2012}.


\begin{figure}
  \centering %
  \mybox{\includegraphics{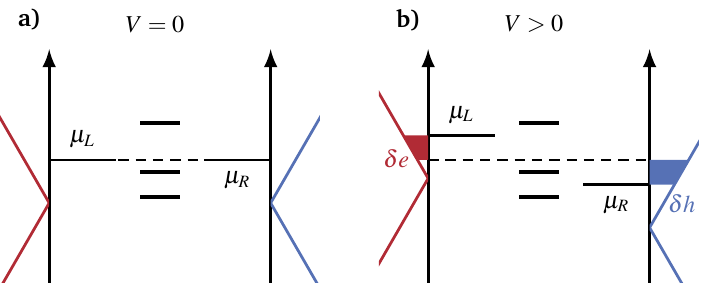}}
  \caption{Illustration of non-symmetric coupling induced by doping out of
      symmetry. \subref a shows the zero bias configuration with broken $e$--$h$ symmetry due to
      doping, $g<0$. \subref b shows a difference among the electrode
      contributions in the bias window which \emph{pins} the system to the right electrode.
      \label{fig:mol-state}
  }
\end{figure}

\paragraph*{Voltage drop model.}%
%
We will now consider a simple model which can explain the electrode selectivity of the
voltage drop depending on the doping/electrostatic gating. \fref{fig:mol-state} is a
guided reference for the following discussion. The position of the voltage drop can be
obtained by considering the change in charge in the scattering region when applying a
bias. If the scattering region becomes more positive, one can view it as the positive
electrode extending into the scattering region and thus the voltage drop will occur closer
to the negative electrode and \emph{vice versa}. The change in charge in the scattering
region is linked to the change in injected charge from left and right electrodes in the
bias window, as noted in the Methods section. The linear dependence of the DOS in the
graphene electrodes makes the coupling/broadening functions of the scattering region
display the energy dependence,
\begin{equation} 
  \label{eq.Gamma}
  \Gam_{L/R}(E)\propto |E-\mu_{L/R}+E_F|,
\end{equation}
where $E=0$ corresponds to the equilibrium Fermi level, $E_F$ is the shift of Fermi level due to doping, 
$E_F\propto\sqrt{g}$, and $\mu_{L/R}$ is the change in the chemical potentials of left/right electrodes with applied
voltage bias ($V$). We will use $\mu_L=eV/2$ and $\mu_R=-eV/2$, and take $V>0$. This
definition means that the scattering region as a starting point will not preferentially
select the left or right electrode for an electron-hole symmetric system, and the
potential drop profile will be spatially anti-symmetric, $\Delta V(x)=-\Delta V(-x)$. We
will now consider the voltage bias as a ``perturbation'' onto the system without bias, and
calculate the change in charge in the scattering region. Thus we first neglect the change
in potential set up by the change in charge, which again will impact the charge in the
self-consistency. With this we have the density of scattering states from left and right,
$A_{L/R}\propto \Gam_{L/R}\propto |E-\mu_{L/R}+E_F|$, and the change in electrons(holes)
injected from left(right) electrode can be written as,
\begin{align}
  \label{eq:drho:L}
  \delta e&=\int_{0}^{eV/2}\!\!\!\!\!A_L\big(E\big)dE
  \propto \tfrac{eV}2\big(E_F-\tfrac{eV}4\big),
  \\
  \label{eq:drho:R}
  \delta h&=\int_{-eV/2}^{0}\!\!\!\!\!A_R\big(E\big)dE
  \propto \tfrac{eV}2\big(E_F+\tfrac{eV}4\big)
\end{align}
where we assume $|V/2|<|E_F|$. The scenario is shown schematically in \fref{fig:mol-state}b
showing more injection of positive carriers $\delta h > \delta e$. Thus the scattering
region will as the first response to the nonequilibrium filling become more positive and
we conclude that for \emph{n}-doping, $g<0$ and $E_F>0$, the positive electrode will
``extend'' into the constriction resulting in a voltage drop at the negative electrode, as
seen in \fref{fig:pot-drop}. We stress that this behavior stems from the vanishing DOS of
graphene at the Dirac point yielding a large relative difference between the electron/hole
contributions. Contrary if we take $E_F$ to be very large in Eqs.~\eqref{eq:drho:L} and
\eqref{eq:drho:R} we get $\delta e\approx\delta h$ and the constriction does not change
its charge. Indeed, the pinning effect is smaller at $\unitr{1}V$ compared to
$\unitr{0.5}V$ as seen in \fref{fig:pot-drop}a vs. \subref c. This is due to the DOS of one lead being very close to zero at $\unitr{0.5}V$; $\mu_i-E_F\approx E_D$ with $E_D$ being
the Dirac point, and hence a much larger relative difference in DOS.

\begin{figure}
  \centering%
  \mybox{\includegraphics[width=.7\columnwidth]{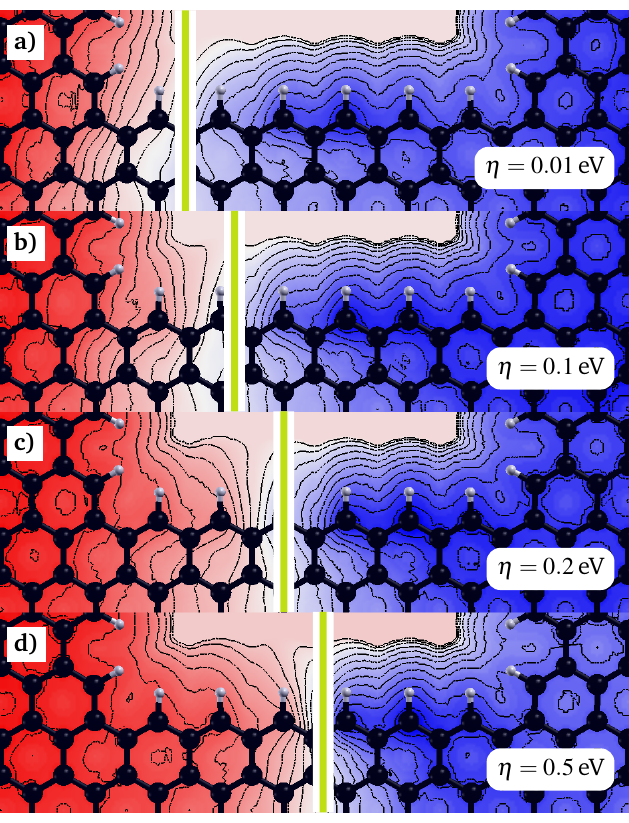}}
  \caption{Change of potential drop vs. level broadening parameter, $\eta_{L/R}$, for
      $\unitr{0.5}V$. Increasing values smear out the electrode DOS which evens out the electronic
      contribution from both electrodes in the bias window. The voltage drop becomes
      anti-symmetric at even charge injection rates from the two electrodes (large smearing).
      \label{fig:eta-smear}
  }
\end{figure}

In order to substantiate that the voltage drop is controlled by the vanishing electrode
DOS we smear the DOS energy dependence gradually into a flat function by introducing an
artificial increase in the broadening parameter, $\eta$, for the electrode self-energies
in \eref{eq:negf}. Hence $\Gam_L(E)\approx \Gam_R(E)$ for $\eta\gg0$ irrespective of the
applied bias and gating. This forces $\delta e\approx\delta h$ and a resulting
anti-symmetric voltage drop. \fref{fig:eta-smear} shows the voltage drop in the middle
part of the constriction for four $\eta$ values. It is clear the anti-symmetric voltage
drop is regained when $\eta_{L,R}\ge\unitr{0.5}{eV}$.  Note that since we have not made
assumptions in the model about the nature of the constriction we anticipate that it can
straightforwardly be applied to similar systems between graphene electrodes in the
high-conductance regime. 


\paragraph*{Constriction, hydrogen terminated with dopants.} %

Since Gr consists entirely of surface atoms it is also extraordinarily susceptible to
external influences such as chemical modification or charged impurities. We will now
discuss the influence of modifying the passivation or having
adatoms\cite{nakada_migration_2011,marconcini_atomistic_2012,Lopez-Bezanilla2009,biel_anomalous_2009}
as a source of charge doping alternative to the electrostatic gating. We examine the
effect of a donating Lithium (Li) or an accepting Flourine (F) adatom placed either inside or
outside the constriction at the positions shown in \fref{fig:bias-drop-H}a. The Li or F
atoms are positioned above the center of a hexagon, or ontop a Carbon atom,
respectively. In \fref{fig:constriction-both}c we show the transmission for the different
adatom configurations. The transmission spectra indicate that very little scattering due
to the dopants themselves takes place, especially when the adatoms are positioned outside
the constriction. The doping effect is clearly seen from the shift in the two resonance
peak positions. Li will \emph{n}-dope the graphene constriction while F \emph{p}-dope
it. Surprisingly, we find that most of the charge transfer to the device resonances is
maintained when the dopants are moved outside the constriction. This suggests that
nanostructured graphene devices will not necessarily be very sensitive to the actual
position of the adatoms. In the case of F it is actually more efficient outside the
constriction. Comparing the most significant peak with the field effect gating
transmission curves we find that Li donates at least $0.2$ electrons while F accepts at
least $0.3$ electrons from graphene. In addition, we find that a pinning of the potential
to the positive/negative electrode occurs for Li(\emph{n}-doping)/F(\emph{p}-doping) for
positive bias, consistent with the potential drops obtained from field effect gating (see
\fref{fig:bias-drop-H}). Adatoms may therefore provide an alternative way to manipulate the
voltage drop by pinning the potential to either of the two electrodes. This underlines the
conclusion that the main effect is determined by the addition or removal of charge from
the device, together with the uneven injection rates from the electrodes.

\section{Conclusion}
\label{sec:conclusion}

We have implemented an electrostatic gate method which introduce charge carriers and the
corresponding electric field in a capacitor-like setup in self-consistent DFT-NEGF
calculations with open boundary conditions to semi-infinite electrodes. The gate method
has been applied to several graphene constrictions where the narrowest junction
corresponds to a graphene nanoribbon with either Hydrogen or Oxygen passivation. For
positive voltage bias and with electrostatic gating the junction potential gets
preferentially pinned to the positive(negative) electrode for $n$($p$)-type doping charge,
and \emph{vice versa} for polarity changes of gating and/or bias. Thus the position of the
voltage drop can be manipulated by the gate potential or correspondingly from charge
doping from adatoms. The constrictions was found to couple selectively to the electrode
with the highest DOS contribution in the bias window. The behavior was traced back to the
vanishing DOS of graphene close to the Dirac point. A simple perturbation model showed how
the selectivity is due to the low DOS of graphene around the Fermi level, irrespective of
the details of the junction electronic structure. The V-shaped DOS is also true for the
local DOS at armchair edges\cite{Ryndyk2012}. Thus we anticipate that our results also
apply to molecular junctions more weakly coupled via a barrier to armchair edges of
graphene. We suggest that this selectivity and high potential gradient can be utilized in
experiments on nanostructured graphene or similar 2D materials to control regions of
reactivity, manipulate polar adsorbates, or providing control of and insights into the
local Joule heating \cite{Dong2007,Murali2009}. We expect that Kelvin Atomic Force
Microscopy\cite{Yan2011}, Scanning Tunnelling Potentiometry\cite{Willke2015} or Low-Energy
Electron Potentiometry\cite{Kautz2015} to be suitable experimental techniques to
examine the effect pointed out here in nanostructured graphene.

\section*{Methods}
\label{sec:theory}

The simulations have been performed using the SIESTA/TranSIESTA code with the PBE-GGA
functional for exchange-correlation\cite{PeBuEr.96} and a SZP basis-set. A confinement
radii determined from an energy shift of $\unitr{230}{meV}$. The real-space grid cutoff
was $\unitr{230}{Ry}$. The electronic temperature has been set to $\unitr{25}{meV}$
($\unitr{50}{meV}$ for the O-terminated constriction). Unless stated otherwise, the
smearing parameter $\eta$ was set to $\unitr{10^{-2}}{eV}$. The geometries were relaxed
until all forces were smaller than $\unitr{5\times10^{-2}}{eV/\Ang}$. Five transverse
$k$-points were used in the electronic structure calculation. This was increased to
between $25$ and $50$ $k$-points in the transport calculations. The transmission data have
subsequently been interpolated\cite{Falkenberg2015}. A vacuum gap of $\unitr{120}{\Ang}$
was used in the direction normal to the constriction plane.

Our field effect setup consists of a gate electrode, a dielectric, and the system, here
being the graphene nanojunctions. Applying a gate voltage charges the system and
electrodes like in a capacitor setup, thus inducing an electrostatic potential gradient
across the dielectric, which in this implementation is \emph{vacuum}. The additional
charge will redistribute to create a polarization in the system along the electric field
direction. Such field effect setups can be realized in open-boundary DFT calculations by
employing a nonequilibrium Green function (NEGF) scheme\cite{Brandbyge2002,Areshkin2010},
or by solving the Poisson equation with appropriate boundary
conditions\cite{Bengtsson1999,Otani2006}. The former is a computationally expensive
calculation compared to the latter.

Analogous to a plate capacitor setup we assume that an applied gate voltage induces an
electron charge $-\delta e^-$ in the system and a corresponding counter-charge $+\delta
e^-$ in the gate plane. This situation is accounted for by charging the system with a
given electron charge $g= -\delta e^-$, and by distributing homogeneously the
corresponding counter-charge $+\delta e^-$ in a well defined region of the unit-cell,
denoted \emph{gate}, so that the overall system+\emph{gate} remains charge neutral. The
setup is shown schematically in \fref{fig:impl-6x}a. Thus for $g>0$ we have a
\emph{p}-doped system, similarly for $g<0$ we have a \emph{n}-doped system. Solving the
Poisson equation inherently calculates the electric field between the gate and the
system. As the calculation cell is periodic we apply the slab dipole
correction\cite{Bengtsson1999} to terminate the periodic electric field induced by the
charge redistribution. The gating method can readily be adopted to transport calculations
using NEGF if the gate is uniformly applied to the electrodes \emph{and} the
device. Additionally, the gate at the electrodes must have a resulting electric field
perpendicular to the applied bias to assert the correct boundary conditions. Our
implementation resembles that of Brumme \etal\cite{Brumme2014,Brumme2015} except that we
use a linear combination of atomic orbitals method, which means that the dielectric need
not be simulated by a potential barrier to limit electronic penetration.

We note that the DFT-NEGF\cite{Brandbyge2002} calculation relies on calculating the
density by occupying the left and right scattering states to the different respective
chemical potentials. This is done by integrating the left/right spectral density matrices,
$\Spec_{L/R}$, given in terms of the retarded Greens function, $\Gf$,
\begin{gather}
  \label{eq:negf}
   \Spec_{L/R}(E) = \Gf(E)\Gam_{L/R}(E)\Gf^\dagger(E),
  \\
 \Gf(E) = \big[(E+i\eta)\Over - \Ham - \SE_L(E) - \SE_R(E)\big]^{-1}.
\end{gather}
Here $\Ham$, $\Over$, $\Gam_{L/R}(E)=i[\SE_{L/R}(E)-\SE_{L/R}^\dagger(E)]$ are the
Hamiltonian, the overlap and the electrode broadening matrices. The parameter $\eta\to0^+$
introduce a vanishingly small broadening of DOS. However, a finite $\eta$
broadens the electrode DOS.

The simple \emph{Voltage drop model} is developed based on the following more detailed
description. We consider a left-right symmetric conductor. In nonequilibrium the density
(matrix) can formally be written at as an ``equilibrium'' contribution corresponding to
the equilibrium Fermi energy, $E_F$, plus two ``nonequilibrium'' contributions originating
from the change in filling of left and right originating scattering states, say, $\mu_L >
E_F >\mu_R$). The ``nonequilibrium'' terms corresponding to negative charge injection from
the negative electrode, and positive charge injection from the positive electrode,
\begin{align}
  \label{eq:negf:A}
  \boldsymbol\rho=& 
  -\frac1\pi\int dE\, \mathrm{Im}\Gf(E)\, n_{F,E_F} +
  \delta e - \delta h,
\end{align}
where $\delta e$ and $\delta h$ are defined in Eqs.~\eqref{eq:drho:L} and \eqref{eq:drho:R}.
We choose $E_F=(\mu_L+\mu_R)/2$ and consider
the different fillings as a perturbation. If we neglect the resulting Landauer dipole field
in $\Ham$, which appear in the response to this perturbation in the self-consistent
DFT-NEGF calculation, then the first ``equilibrium'' term can not break left-right
symmetry and result in a left-right symmetric density. It is then clear that the symmetry
breaking and charge in the device is determined by the competition between the latter two
contributions which are of opposite sign.


The systems studied here belong to the class highly conducting carbon junctions for which
the DFT-NEGF method has been compared favorably to detailed experiments both in the
linear\cite{Schull2009,Frederiksen2014} and non-linear conductance regime
\cite{Schneider2015}. In any case, we are here mainly interested in the qualitative
aspects of the behavior of the voltage drop.

\section*{Acknowledgements}

We thank Osamu Sugino and Pablo Ordej\'on for useful comments on the method employed,
and the Danish e-Infrastructure Cooperation (DeIC) for providing computer resources.
The Center for Nanostructured Graphene (CNG) is sponsored by the Danish Research
Foundation, Project DNRF58. DS acknowledges the support from the H.C. {\O}rsted COFUND
postdoctoral program at DTU.

\nocite{Kokalj1999}

\end{document}